\documentclass{article}
\usepackage{amsmath}
\usepackage{amssymb}

\input{tcilatex}

\begin{document}

\begin{center}
\bigskip \textbf{SCALAR FIELD ENTROPY IN\ BRANE WORLD\ BLACK\ HOLES}

\bigskip

\bigskip

\bigskip

K.K. Nandi$^{1,2,a}$, Y.-Z. Zhang$^{2,b}$, A. Bhadra$^{4,c}$ \ \

and

P. Mitra$^{5,d}$

$\bigskip $

$\bigskip $

$^{1}$Department of Mathematics, University of North Bengal, Darjeeling
(W.B.) 734 430, India


$^{2}$Institute of Theoretical Physics, Chinese Academy of
Sciences, P.O.Box 2735, Beijing 100080, China

$^{4}$IRC, University of North Bengal, Darjeeling (W.B.) 734 430, India

$^{5}$Theoretical Nuclear Physics Division, Saha Institute of Nuclear
Physics, 1/AF Bidhan Nagar, Salt Lake, Calcutta 700 064, India

--------------------------------------------------------------

\bigskip
\end{center}

$\bigskip $

$\bigskip $

$^{a}$E-mail address: kamalnandi1952@yahoo.co.in

$^{b}$E-mail address: yzhang@itp.ac.cn

$\bigskip ^{c}$E-mail address: aru\_bhadra@yahoo.com

$^{d}$E-mail address: parthasarathi.mitra@saha.ac.in

\begin{center}
\bigskip \textbf{Abstract}
\end{center}

A semiclassical calculation of entropy of a scalar field in the background
of a class of brane world black holes (BWBH) is carried out in the presence
of a brick wall cutoff when the 5D-bulk induced \textquotedblleft tidal
charge" has generic or extreme values. 

\begin{center}
------------------------------------
\end{center}

The entropy $S$ of a standard nonextremal black hole obeys the well known
Bekenstein-Hawking area law $S=\frac{A}{4G}$, where $A$ is the area of the
horizon. This result can be rederived by the Lagrangian path integral
formalism [1,2] or by other means [3]. However, the entropy due to quantum
fields in the black hole background introduces divergences which are
interpreted as renormalizations of the gravitational coupling constant $G$.
In the extreme case defined by two coinciding surfaces, the situation is
different - the area law does not hold ($S=0$). It has been argued that
extreme and nonextreme BHs should be regarded as qualitatively different
objects due to discontinuity in the Euclidean topology [4]. In this context,
one would recall that dilatonic black holes [5] have zero area at
extremality and hence $S=0$. However, $S$ does not remain zero when quantum
contributions are taken into account: The linear divergence goes away but a
nonleading logarithmic divergence persists [6].

Given this scenario and given the current widespread interest in the brane
theory, it would be important to assess similar quantum contributions to
entropy in the background of BWBHs. The purpose of the present note is to do
this. The motivation is that, generically, the BWBHs are different from
ordinary BHs as they embody a synthesis of wormhole and blackhole features.
For instance, the effective stress energy tensor violates some of the energy
conditions. As argued in Refs.[7,8], this feature is not unexpected as the
tensor contains imprints of the \textit{free} gravitational field in the
bulk which contributes negative energy. Essentially, the bulk contribution
implies a correction to the Schwarzschild solution but its horizon structure
remains undisturbed. The brane theory we have in mind is described by the
RS2 framework, that is, a single brane in a $Z_{2}$-symmetric 5-D
asymptotically anti-de Sitter bulk in which only gravity propagates while
all other fields are confined to the brane [9]. The 5-D Weyl tensor
projected onto the brane is represented by a trace-free tensor $E_{\mu
}^{\nu }$ appearing in the Shiromizu-Maeda-Sasaki equations [10]. The trace
of these equations, viz., Ricci scalar $R=0$, are solved to derive the
BWBHs. Henceforth, we take units $G=c=\hslash =k=1$, and Greek indices run
from $0$ to $3$.

The general class of 4-D solutions is given by [11]

\begin{equation}
ds^{2}=g_{\mu \nu }dx^{\mu }dx^{\nu }=-A(r)dt^{2}+\frac{dr^{2}}{B(r)}%
+r^{2}d\Omega ^{2}
\end{equation}%
in which $d\Omega ^{2}$ is the metric on the unit $2$-sphere, $A(r)$ and $%
B(r)$ are two well behaved positive functions for $r>r_{h}$ and have a
simple zero at $r=r_{h}$ defining the horizon. The singularities, if any, of
the BH solutions when propagated off the brane into the 5-D bulk may make
the AdS horizon singular (\textquotedblleft black cigar" [12]). However,
several classes of nonsingular, static, spherically symmetric BWBH solutions
have now been proposed almost simultaneously [13-15] and some quantum
properties have also been investigated [16]. Under certain assumptions on
the behavior of the metric functions, Bronnikov, Melnikov and Dehnen [11]
have shown that the solution (1) can have $R\times S^{2}$ topology of
spatial sections. Assuming asymptotic flatness at large $r$, the global
causal structure of such solutions coincides with that of Kerr-Newman
nonextremal solutions. We shall specifically consider a class of solutions
given by [11,13]:

\begin{equation}
A(r)=1-\frac{2M}{r};B(r)=\frac{(1-\frac{2M}{r})(1-\frac{r_{0}}{r})}{1-\frac{%
3M}{2r}}
\end{equation}%
where $M$ and $r_{0}(\neq 0)$ are two adjustable parameters, one is related
to the mass and the other may be interpreted as an induced tidal charge\ - a
Weyl tensor projection from the bulk into the brane. The ranges $%
r_{0}>2M,r_{0}=2M$ and $r_{0}<2M$ correspond respectively to traversable
wormholes, extremal and nonextremal regular BWBHs. The areas of both
categories of black holes are nonzero. For $r_{0}=3M/2$, one recovers the
Schwarzschild solution.

Let us consider the solution (2) and calculate the background values of $S$
and $\kappa $ following
[3]:
\begin{equation}
S=-\frac{1}{8\pi }\int_{Hor}[K]=\frac{A(r_{h})}{4},
\end{equation}%
where $[K]=K-K_{0}$ in which $K$ is the extrinsic curvature and $K_{0}$ is
obtained by substituting into $K$ the flat space metric. The surface gravity
$\kappa $ at the horizon $r_{h}=2M$ is given by

\begin{equation}
\frac{\kappa }{2\pi }=\frac{1}{\beta }=\frac{1}{4\pi }\frac{\partial
_{r}g_{00}}{\sqrt{-g_{rr}g_{tt}}}=\frac{1}{4\pi M}\left( 1-\frac{r_{0}}{2M}%
\right) ^{\frac{1}{2}}.
\end{equation}

We shall now adopt the statistical mechanics approach via the partition
function [2,6,17] for the quantization of a scalar field field $\phi $ in
the fixed background given by the solution (1). The dominant contribution to
the partition function $Z$ will be assumed to come from the classical
solutions of the Euclidean Lagrangian action $S_{1}[g_{cl}]$ leading to the
area law in general. Neglecting the quantum fluctuations of the metric and
other bulk induced quantities, that is, freezing them into their background
values, the partition function can be represented as

\begin{equation}
Z=e^{-S_{1}[g_{cl}]}\int [d\phi ]e^{-S_{2}[g_{cl},\phi ]}.
\end{equation}%
In order to find the contribution of $\phi $ to the partition function $Z$
through $S_{2}$, we further employ the \textquotedblleft brick
wall\textquotedblright\ boundary condition of 't Hooft [17]. That means, we
take $\phi (r)=0$ at $r=2M+\epsilon $ where the small quantity $\epsilon $
signifies an ultraviolet cut-off. There is also an infrared cut-off, that
is, $\phi (r)=0$ at $r=L$ where $L>>2M$. The Klein-Gordon equation for this
scalar field in the background of the spacetime $g_{\mu \nu }$ reads

\begin{equation}
\frac{1}{\sqrt{-g}}\partial _{\mu }\left( \sqrt{-g}g^{\mu \nu }\partial
_{\nu }\phi \right) -m^{2}\phi =0.
\end{equation}%
An ansatz of the form $\phi =e^{-iEt}f_{El}Y_{lm}$ yields the radial
equation for the metric (1):

\begin{equation}
A^{-1}E^{2}f_{El}+\left( \frac{B}{A}\right) ^{\frac{1}{2}} \frac{1}{r^2}%
\frac{\partial }{\partial r}\left[ \left( AB\right) ^{\frac{1}{2}}r^{2}\frac{%
\partial f_{El}}{\partial r}\right] -\left[ l(l+1)+m^{2}\right] f_{El}=0.
\end{equation}%
Taking further $f_{El}\sim e^{i\int dr p}$, and keeping the real parts, a
radial wave number can be found as

\begin{equation}
p^{2}=B^{-1}\left[ A^{-1}E^{2}-l(l+1)-m^{2}\right] .
\end{equation}%
Restricting $p$ to the real values, we introduce the semiclassical
quantization condition

\begin{equation}
\pi n_{r}=\int_{2M+\epsilon }^{L}p(r,l,E;m)dr
\end{equation}%
where $n_{r}$ must be a nonnegative positive integer. Then, with $\beta
=1/T=2\pi /\kappa $, where $T$ is the temperature, $\kappa $ is the surface
gravity, the Helmholtz free energy $F$, appearing in $Z=e^{-\beta F}$, is
given by

\begin{eqnarray}
\beta F &=&\sum_{n_{r},l,m}\ln (1-e^{-\beta E})\approx \int dl(2l+1)\int \ln
(1-e^{-\beta E})dn_{r}  \notag \\
&=&-\int dl(2l+1)\int d(\beta E)\left( e^{\beta E}-1\right) ^{-1}n_{r}
\notag \\
&=&-\frac{2\beta }{3\pi }\int_{2M+\epsilon
}^{L}(A)^{-3/2}(B)^{-1/2}r^{2}dr\int dE\left( e^{\beta E}-1\right) ^{-1}%
\left[ E^{2}-Am^{2}\right] ^{\frac{3}{2}},
\end{eqnarray}%
where the $l$ integration has been explicitly carried out.

Now, as is customary [6], we set the lower limit in the $E$ integral to near
zero and evaluate it approximately ignoring the contributions coming from
the $m^{2}\beta ^{2}$ term. Then we find the following result for $M\neq 0$
from Eq.(8): There is a linearly divergent contribution for $\epsilon
\rightarrow 0$ given by

\begin{equation}
F_{corr}\approx -\frac{4\pi ^{3}}{180\epsilon }\left( \frac{2M}{\beta }%
\right) ^{4}\left( 1-\frac{r_{0}}{2M}\right) ^{-1/2}.
\end{equation}%
%
%
%
%
%
%
The entropy $S$ can be expressed in terms of the free energy as

\begin{equation}
S=\beta ^{2}\frac{\partial F}{\partial \beta }
\end{equation}%
and the contribution from the linearly divergent term is given by

\begin{equation}
S_{corr}=\frac{16\pi ^{3}}{180}\left( \frac{2M}{\beta }\right) ^{3}\frac{%
\left( 1-\frac{r_{0}}{2M}\right) ^{-1/2}}{2M\epsilon }(2M)^{2}.
\end{equation}%
%
%
%
%
%
%
At the horizon, on putting the value of $\beta $ from (4), this reduces to

\begin{equation}
S_{corr}=\frac{1}{90}\frac{\left( 1-\frac{r_{0}}{2M}\right) }{2M\epsilon }%
(2M)^{2}.
\end{equation}%
Thus the nonextremal ($r_{0}\neq 2M$) behavior of the cut-off dependence ($%
\epsilon \rightarrow 0$) is similar to that in other black holes.

It is important to introduce the proper thickness $\tilde{\epsilon}$ of the
black hole at this point:
\begin{equation}
\tilde{\epsilon}=\int_{2M}^{2M+\epsilon }\frac{dr\sqrt{1-\frac{3M}{2r}}}{%
\sqrt{1-\frac{2M}{r}}\sqrt{1-\frac{r_{0}}{r}}}=\frac{\sqrt{2M\epsilon }}{%
\sqrt{1-\frac{r_{0}}{2M}}}.
\end{equation}%
The entropy $S_{corr}$ can be rewritten as
\begin{equation}
S_{corr}=\frac{1}{90}\frac{ 1 }{\tilde{\epsilon}^{2}}(2M)^{2}.
\end{equation}

Thus, if we keep $\tilde{\epsilon}$ fixed, we find that, in the extremal
limit $r_{0}\rightarrow 2M$, $S_{corr}$ remains finite. This is so in spite
of the vanishing of Hawking temperature in this limit (Actually the factor $%
(1-\frac{r_{0}}{2M})$ cancels out). The reason for this is that the
divergence in the $r$ integral becomes stronger and compensates the
temperature effect.
The physical difference between extreme and near extreme BHs have been
discussed at length in Ref.[4].

A direct calculation may be carried out for the extremal case. In this
situation, the expression for the temperature given above vanishes, but the
Euclidean black hole has no conical singularity and the time can be given
any periodicity, so that the temperature is \textit{arbitrary}. With this
understanding, one can write the contribution from the field to the entropy
as
\begin{equation}
S_{corr}^{ex}=\frac{8}{135}\pi ^{3}\left( \frac{2M}{\beta }\right)
^{3}\left( \frac{2M}{\epsilon }\right) ^{3/2}=\frac{8}{135}\pi ^{3}\left(
\frac{2M}{\beta }\right) ^{3}\exp (\frac{3}{2}\Lambda ),
\end{equation}%
where $\Lambda $ is a \textit{large} cutoff representing the proper distance
from the horizon (at negative infinity in $r$) to the brick wall [18]. This
expression shows much stronger exponential divergence, as compared to the
quadratic divergence shown in the extreme limit of the non-extremal case.

The above results for the BWBH may be compared with the thermodynamics of
the Reissner-Nordstr\"{o}m solution. In both cases, the period $\beta $ in
Euclidean time is infinite in the extremal limit and arbitrary at exact
extremality, giving zero entropy for the purely gravitational theory [4]
(see Refs. [19,20] for a different approach to quantization). The
nonextremal contributions of the matter field contain mild divergences, but
the extremal limit does not produce anything stronger: $S_{corr}$ goes to a
finite limit. The extremal black holes, if treated directly, show a stronger
divergence. In the dilatonic solution, the area law holds [3,21] and in the
extremal limit or extremal case the leading contribution to $S_{corr}$
disappears since $A=0$, but leaves a logarithmic divergence [6].

We can extend the approach in this paper to another type of BWBH for which
the metric functions are [7,8,13-15]

\begin{equation}
A(r)=B(r)=1-\frac{2M}{r}-\frac{Q^{2}}{r^{2}},
\end{equation}%
in which $M$ is the mass and $Q$ is the tidal charge. Due to the \textit{%
wrong} sign before $Q^{2}>0$, the stress energy violates the energy
conditions. It should be remarked that, historically, this kind of solution
was \textit{heuristically} conceived by Einstein and Rosen in their attempt
to construct singularity free particle models (\textquotedblleft
Einstein-Rosen bridge") [22]. However, the brane world physics of today
provides a physical justification for this sign reveresed solution via tidal
effects \textit{albeit} the spacetime shares some properties of a wormhole
(two sheeted topology, for instance). Nevertheless, the solution is treated
as a BWBH and is recently applied in the context of gravitational lensing
[23]. The BH horizons appear at $r_{h}^{\pm }=M\pm \sqrt{M^{2}+Q^{2}}$ and
the temperature of the outer horizon $r_{h}^{+}$ \ is given by

\begin{equation}
T=\frac{\sqrt{M^{2}+Q^{2}}}{2\pi \left[ M+\sqrt{M^{2}+Q^{2}}\right] ^{2}}
\end{equation}%
and the area law is satisfied. But, $T=0$ is possible only if $M=0$ and $Q=0$%
, that means, a flat space. Thus, nontrivial extreme solution like the ones
dealt with above simply does not exist here! Calculations show that there
occur both linear and logarithmic divergences including in the case $M=0$
(pure tidal charge solution). We withhold the details except only mentioning
that these divergences persist also in all other types of BWBHs proposed in
Ref.[11]. In this sense, the first example occupies a very special place
indded.

To summarize, we have brought out the quantum behavior of a scalar field in
the background of BWBHs. It is found that nonextremal BWBHs produce
ultraviolet ($\epsilon \rightarrow 0$) divergences in entropy, as do other
ordinary BHs. The nontrivial extremal BWBH produces stronger (exponential)
divergences like in the RN extreme case. This is perhaps understandable
noting that both BWBH and RNBH satisfy the same field equation $R=0$, where $%
R$ is the Ricci scalar. However, the extreme counterpart does not exist in
all classes of BWBH with tidal charge. These results underline the
distinctive features of BWBH thermodymamics.

\bigskip

\textbf{ACKNOWLEDGMENT}

\bigskip Part of the work was carried out while one of the authors (KKN) was
visiting Ufa, Russia. It is a pleasure to thank Guzel Kutdusova
for several useful assistance. This project was in part supported
by NNSFC under Grant No. 90403032 and also by National Basic
Research Program of China under Grant No. 2003CB716300.

\textbf{\bigskip }

\textbf{REFERENCES}

\bigskip

[1] G.W. Gibbons and S.W. Hawking, Phys. Rev. D \textbf{15}, 2752 (1977).

[2] L. Susskind and J. Uglum, Phys. Rev. D \textbf{50}, 2700 (1994).

[3] R. Kallosh, T. Ort\'{\i}n, and A. Peet, Phys. Rev. D \textbf{47}, 5400
(1993). For a recent discussion in a general context, see: T. Padmanabhan,
Mod. Phys. Lett. A \textbf{19}, 2637 (2004); Class. Quant. Grav. \textbf{21}%
, 4485 (2004).

[4] S.W. Hawking, G.T. Horowitz, and S.F. Ross, Phys. Rev. D \textbf{51},
4302 (1995).

[5] G.W. Gibbons and K. Maeda, Nucl. Phys. B \textbf{298}, 741 (1988); D.
Garfinkle, G. Horowitz, and A. Strominger, Phys. Rev. D \textbf{45}, 3888
(1992).

[6] A. Ghosh and P. Mitra, Phys. Rev. Lett. \textbf{19}, 2521 (1994).

[7] N. Dadhich, Phys. Lett. B \textbf{492}, 357 (2000); N. Dadhich, R.
Maartens, P. Papadopoulos, and V. Rezania, Phys. Lett. B \textbf{487}, 1
(2000).

[8] S. Shankarnarayanan and N. Dadhich, Int. J. Mod. Phys. D \textbf{13},
1095 (2004).

[9] L. Randall and R. Sundrum, Phys. Rev. Lett. \textbf{83}, 4690 (1999).

[10] T. Shiromizu, K. Maeda, and M. Sasaki, Phys. Rev. D \textbf{62}, 024012
(2000).

[11] K.A. Bronnikov, V.N. Melnikov and H. Dehnen, Phys. Rev. D \textbf{68},
024025 (2003).

[12] A. Chamblin, S.W. Hawking, and H.S. Reall, Phys. Rev. D \textbf{61},
065007 (2000).

[13] R. Casadio, A. Fabbri, and L. Mazzacurati, Phys. Rev. D \textbf{65},
084040 (2002). See also Refs.[7,8,11].

[14] N. Dadhich, S. Kar, S. Mukherjee, and M. Visser, Phys. Rev. D \textbf{65%
}, 064004 (2002).

[15] Exterior metric similar to Eq.(19), but applicable for a star in the
brane, has been derived by: C. Germani and R. Maartens, Phys. Rev. D.
\textbf{64}, 124010 (2001). They also derived the interior metric.

[16] R. Casadio and B. Harms, Phys. Lett. B \textbf{487}, 209 (2000); R.
Casadio, Ann. Phys. (NY) \textbf{307}, 195 (2003); Phys. Rev. D \textbf{69},
084025 (2004).

[17] G. 't Hooft, Nucl. Phys. B \textbf{256}, 727 (1985).

[18] A. Ghosh and P. Mitra, Phys. Lett. B \textbf{357}, 295 (1995).

[19] A. Ghosh and P. Mitra, Phys. Rev. Lett. \textbf{78}, 1858 (1997).

[20] For a grand canonical ensemble approach involving charged black holes,
see: H.W. Braden, J.D. Brown, B.F. Whiting, and J.W. York, Phys. Rev. D
\textbf{42}, 3376 (1990).

[21] J. Preskill, P. Schwarz, A. Shapere, S. Trivedi, and F. Wilczek, Mod.
Phys. Lett. A \textbf{6}, 2353 (1991); C. Holzhey and F. Wilczek, Nucl.
Phys. B \textbf{380}, 447 (1992); R. Kallosh, A. Linde, T. Ort\'{\i}n, and
A. van Proeyen, Phys. Rev. D \textbf{46}, 5278 (1992).

[22] A. Einstein and N. Rosen, Phys. Rev. \textbf{48}, 73 (1935).

[23] R. Whisker, Phys. Rev. D \textbf{71}, \ 064004 (2005).

\end{document}